# Ice-bridging frustration by self-ejection of single droplets results in superior anti-frosting surfaces

Nicolò G. Di Novo[1,2,*], Alvise Bagolini[2,*], Nicola M. Pugno[1,3,*]

[1] *Laboratory of Bioinspired, Bionic, Nano, Meta, Materials & Mechanics, Department of Civil, Environmental and Mechanical Engineering, University of Trento, Via Mesiano, 77, 38123 Trento, Italy*
[2] *Sensors and Devices Center, Bruno Kessler Fundation, Via Sommarive 18, 38123 Trento*
[3] *School of Engineering and Materials Science, Queen Mary University of London, Mile End Road, London E1 4NS, United Kingdom*
[*]*Corresponding authors*

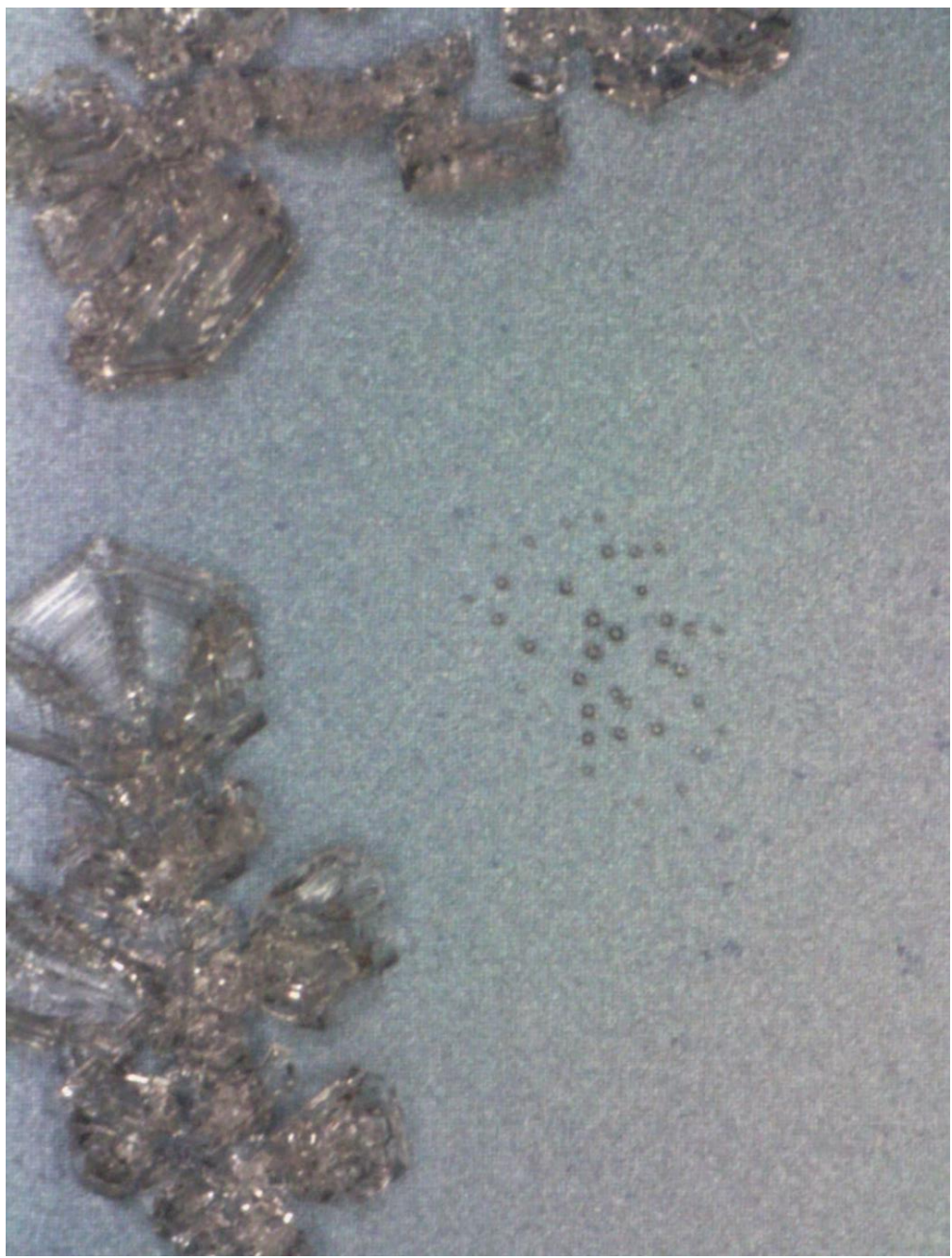

## Abstract

Surfaces capable of delaying the frosting passively and facilitating its removal are highly desirable in fields where ice introduces inefficiencies and risks. Coalescence jumping, enabled by highly hydrophobic surfaces, is already exploited to slow down the frosting but it is insufficient to completely eliminate the propagation by ice-bridging. We show how the self-ejection of single condensation droplets can frustrate the ice bridges of all the condensation droplets leading to a frost velocity lower than 0.5 μm/s thus dropping below the current limits of passive surfaces by a factor of at least 2. Arrays of truncated microcones, covered by uniformly hydrophobic nanostructures, enable individual condensation droplets to growth and self-propel towards the top of the microstructures and to self-eject once a precise volume is reached. The independency of self-ejection on the neighbour droplets allows a precise control on the droplets' size and distance distributions and the ice-bridging frustration. The most performant microstructures tend to cones with a sharp tip on which the percentage of self-ejection is maximum. Looking towards applications, tapered microstructures allow maximising the percentage of self-ejecting drops while maintaining a certain mechanical strength. Further, it is shown that inserted pinning sites are not essential, which greatly facilitates manufacturing.

## Introduction

In agri-food applications, refrigeration systems' efficiency is reduced by the formation of frost and the necessary defrosting cycles[1–4]. In the automotive market, with a growing demand for electric vehicles, the problem of passenger compartment heating in cold climates has arisen, and one solution is in heat pumps whose efficiency is again reduced by the formation of frost[5,6]. Frosting and subsequent ice accretion also affects aircrafts[4,7] and wind turbines[4], changing their aerodynamic profile with a subsequent efficiency reduction. Given the toxic effect of anti-icing and de-icing fluids used in aircrafts[8] and wind turbines, there has been a growing interest in active systems[9] or systems that use waste heat[10] to prevent and remove ice.

In the last decade, research has intensified into passive anti-icing surfaces to complement or potentially replace active anti-icing methods. In particular, the passive anti-icing effect of superhydrophobic surfaces has been studied. These surfaces allow the rebounding and the freezing delay of supercooled droplets due to the reduced solid-liquid contact area and hydrophobic chemistry.[11–14] However, superhydrophobic surfaces exposed to cold and humid environments are not free from the formation of supercooled condensation droplets[15] which may nullify the aforementioned abilities.[16,17]

Moreover, once few droplets freeze by homogeneous or heterogeneous nucleation, the frost propagates through the *ice-bridging* mechanism[18,19] in a process called condensation frosting, ultimately covering the surface with ice, which makes it hydrophilic.[20,21] Regarding the microscopic process, the supercooled condensation droplets evaporate and their vapour desublimates on nearby frozen droplets, being the saturation pressure of liquid water higher compared to that of ice for temperatures lower than 0 °C. This results in the formation of "*ice bridges*" growing from the frozen towards the liquid droplets. What regulates the average frost propagation velocity are the diameter ($d$) and distance ($l$) distributions of the condensation droplets and the surface temperature[15] while the substrate thermal conductivity is negligible in comparison.[22] A liquid droplet that is relatively large and/or close to its frozen neighbour will be reached by the ice bridge, then freeze rapidly and originate the next bridge; this propagation is relatively fast. If, on the other hand, the drop is relatively small and/or far from its frozen neighbour, it will completely evaporate before being reached, causing a slowdown of the bridge which will then be only fed by distant drops. When all the nearest neighbours evaporate, an "ice trail" forms in front of the ice bridges. The more ice trails, the slower the global advancement of the frost. The bridging parameter $S^* \equiv L/d$ gives an indication of the success or failure of ice bridges in contacting nearby liquid drops: for $S^* \lesssim 1$ it is successful and vice versa for $S^* \gtrsim 1$.[23] $L$ is the distance between the liquid droplet centre and the edge of the frozen droplet. The frost velocity is in the order of tens of μm/s for common surfaces with $S^* \lesssim 1$ for all the droplets.

On the basis of this description, introduced by Boreyko et al.,[24] structured surfaces hydrophobic enough to enable coalescence-induced condensation droplet jumping[25–29] were investigated and designed for anti-frosting applications.[30–36] The out-of-plane jumps generate droplet populations with on average smaller diameter and larger spacing than the typical populations of non-jumping coalescing droplets observed on hydrophobic or hydrophilic surfaces. The jumps delay the frost propagation effectively as a high percentage of droplets with $S^* \gtrsim 1$ is maintained over time but, at the same time, being the location of droplets random, a residual percentage of droplets with $S^* \lesssim 1$ establishes the lower limit of the frost velocity around some μm/s. [24,30–36] Alternative anti-frosting strategies are the disposition of the condensation droplets on hydrophilic micropatterns[37] or the use of microstripes that promote ice nucleation and confine frost on them.[38] However, a drawback of such strategies is in the loss of water repellent and anti-icing effects of superhydrophobicity.

We wondered how to realize a superhydrophobic surface capable of suppressing ice-bridging for all the droplets. This aim has guided the conceptualization of single droplet self-ejection from truncated microcones with uniform wettability, theoretically described and experimentally demonstrated in our previous study.[39] The self-ejection makes the jumps independent from the proximity to other drops and happens at a precise volume given by the microstructures' geometry and dynamic contact angles. Moreover, we proofed that inserted micro pinning sites—hydrophilic or less hydrophobic than the surrounding—are superfluous for self-ejection thus facilitating the surface fabrication and scalability. In the present study, we experimentally investigate the anti-frosting effect of the self-ejection phenomenon and how much the head area fraction of the truncated cones ($\varphi_H$) influences the percentage of self-ejecting drops and the frosting speed and coverage. The microstructures tending towards the ideal cone ($\varphi_H \to 0$) are the most performing since almost all the drops self-eject (~90%) and none of the remaining ones satisfies the condition of successful ice-bridging. All the droplets evaporate completely before being touched and the frost front slowly advances purely by evaporation-desublimation forming spectacular crystals that resemble snowflakes. As $\varphi_H$ increases, successful ice-bridging events happen and mix with the pure evaporation-desublimation growth. We here quantify the average speed of the crystals, the global propagation velocity and the frost coverage, and relate them with the truncated microcones' geometry. A frost velocity of 0.4÷1 μm/s is demonstrated in the central portion of the sample, on areas of hundreds/thousands of μm$^2$, while on the full sample scale (1÷4 cm$^2$), where edge effects are influential, it increases to 1÷3.5 μm/s. This novel class of superhydrophobic surfaces establishes a new limit of passive anti-frosting and promises advancements in other applications as condensation heat transfer, water harvesting and self-cleaning.

## Materials and Methods

### Microstructuring

The type of surfaces tested here were presented in our previous study[39] on the self-ejection mechanism. Five surfaces (Group1) served for the fabrication process setup, while the other four (Group2) were employed for characterizing the self-ejection velocity as the droplet radius varies. All consist in arrays of truncated microcones arranged in a square pattern fabricated on 6-inch silicon wafers (100) by photolithography and tapered reactive ion etching (t-RIE), uniformly covered with Aluminium nanostructures (NanoAl) obtained by hot water treatment (HWT) and rendered hydrophobic by silanization (Figure 1).

The following steps are the same for both the groups: after a standard RCA cleaning, the hard mask was made by growing 200 nm of thermal silicon oxide (Centrotherm E1200HT furnace) followed by the deposition of 200 nm of Aluminium by magnetron sputtering (Eclipse MRC). Then we deposited 1.2 μm of positive photoresist by spin coating (Track SVG). We employed two photolithographic masks (Photronics), as described in detail in our previous study:[39] Mask1 for Group1 with patterned areas of 1 cm x 1 cm (mask aligner Suss MA150CC) and Mask2 for Group2 with patterned areas of 2 cm x 10 cm (Nikon stepper model 2205i11D). The masks consist in circles of diameter $D_m$ and pitch $p$ indicated in

Table 1 where each surface is named *Surface_$D_m$ x $p$*. After developing (Track SVG), hard bake of the photo-polymer was carried out. The pattern-transfer onto the hard mask was performed by dry etching of Aluminium (KFT Metal PlasmaPro100 Cobra300) and Silicon oxide (Tegal 903e). Then the t-RIE step was performed (Alcatel dry etcher) with the following parameters: source power 2800 W, bias power 20 W, gas fluxes ratio $SF_6/C_4F_8$=0.65, total gas flux 500 sccm, chamber pressure 0.04 mbar and wafer temperature 20 °C. We removed the etching passivation layer by immersion in isopropanol with ultrasonic pulses.

The hard mask removal was different for the two groups. For Group1, HF vapour (PRIMAXX® uEtch System) was used to etch the silicon oxide under the Al circles, followed by a jet of deionized water to mechanically remove them. As a result, the hard mask is completely removed from the microstructures' top but remains of the flat parts of the wafers. For Group2, we removed the hard mask entirely by immersing the wafers in an Al etch solution and then in a Piranha solution. We cleaned all the wafers in a deionized water rinse until the bath reached 16 MΩ-cm. The truncated microcones have a slight undercut at the apex and we removed it with isotropic etching (Tegal 900) which lowered the pillars by about 2 μm and made the top part straight.

The tapering $\beta$ (reported in Table 1 and obtained with the procedure described in Supplementary information S1) is constant from the bottom base up to about three fourths of the microstructure, then it goes to zero in the upper cone portion. In Table 1 we report all the geometrical parameters measured and calculated by analysing SEM (Vega3, Tescan) images. A parameter that we shall use in the analysis of the performances is the *head area percentage* ($\varphi_H$) defined in Eq. 1, where $d_h$ is the head diameter of the truncated microcone.

$$\varphi_H = \left[\frac{\pi}{4}\left(\frac{d_h}{p}\right)^2\right]100 \tag{1}$$

*Table 1. Geometry parameters and HWT times of the nine tested surfaces.*

| Surface name | $D_m$ [μm] | $p$ [μm] | $\beta$ [°] | $d_h$ [μm] | Height [μm] | $\varphi_H$ [%] | HWT time [s] |
|---|---|---|---|---|---|---|---|
| Group 1 ||||||||
| *5x10* | 5 | 10 | 3.6 | 1.2 | 16.4 | 1.13 | 450 |
| *10x15* | 10 | 15 | 4.9 | 3.9 | 30.7 | 5.31 | 540 |
| *10x20* | 10 | 20 | 3.5 | 3.9 | 29.1 | 2.98 | 540 |
| *15x20bis* | 15 | 20 | 6.2 | 5.8 | 41.1 | 6.61 | 490 |
| *15x25* | 15 | 25 | 4.5 | 5.4 | 42.0 | 3.66 | 490 |
| Group 2 ||||||||
| *10x13* | 10 | 13 | 6.6 | 5 | 23.3 | 11.62 | 420 |
| *15x20* | 15 | 20 | 6.6 | 7 | 34.9 | 9.62 | 420 |
| *30x40* | 30 | 40 | 10 | 12.5 | 64.3 | 7.67 | 420 |
| *60x80* | 60 | 80 | 8.8 | 31.5 | 103.7 | 12.18 | 420 |

## *Nanostructuring, silanization and contact angles*

As a second hierarchical level we employed nanostructured Aluminium (NanoAl) obtained by HWT). The HWT of many metals and their alloys leads to the formation of nanostructures: a thin superficial layer of metal oxide forms in hot water, the oxide cations are released in solution, migrate and deposit forming nanostructures with peculiar shapes for each metal[40]. In the case of Al, thin nano-blades of hydrated aluminium oxide (pseudo-boehmite) form[41] which, once made hydrophobic, gain superhydrophobic and anti-freezing properties.[67–73] We initially deposited 150 nm of pure Al on the microfabricated wafers by e-beam evaporation (ULVAC HIGH VACUUM COATER EBX-16C). The wafers were cleaved in samples of 2 cm x 2 cm for Group2 (in order to have samples with a row of cones on the sharp edge and to observe droplet jumps from side-view) and in samples of ~1.3 cm x 1.3 cm for Group1. HWT was performed by immersion in deionized water (18 MΩ-cm) at 90 °C. As reported,[39] the HWT time of Group2 is 7 min and the darkening, due to nanostructuring, is simultaneously visible on the entire sample and starts a few seconds after immersion. The Group1 samples, as explained, have the patterned area free from the hard mask which remains only on the flat contour part. On preliminary samples processed with the procedure of Group1, we noticed a delay of the darkening during HWT. In particular, it appeared after some tens of seconds, starting from the centre of the patterned area — far from the hard mask — and rapidly propagating towards the edges of the patterned area. This observation offered an insight and a useful guideline for NanoAl fabrication. Two pioneering studies of D. A. Vermilyea and W. Vedder, dating back to 1969-70, report the inhibition of aluminium and water reaction by the presence of other compounds, e.g. silicic acid ($H_4SiO_4$).[41,42] The silicon oxide dissolves in solution and forms silicic acid[43] thus the thin silicon oxide sublayer of the hard mask acts as a retardant of nanostructuring. In order to have a similar nanostructuring between the two groups, we immersed the Group1 samples, waited for the darkening to begin (appreciable to the naked eye) and left 7 minutes to pass before extraction. Table 1 shows a certain variability of the waiting time that depends on the flat area fraction. After HWT, the samples of both groups were immediately immersed in room-temperature deionized water to block the nanostructuring and dried with a $N_2$ flow.

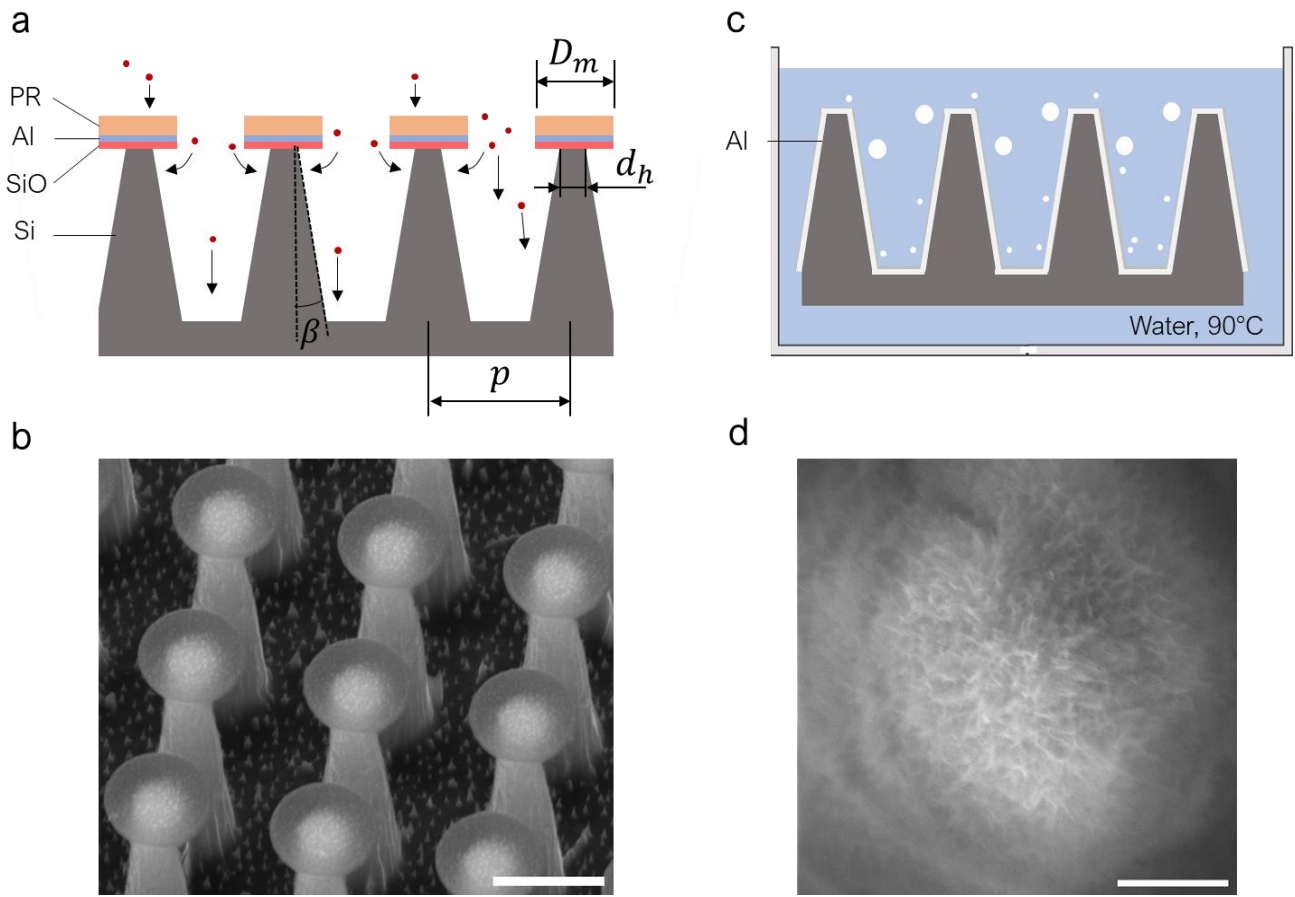

*Figure 1. a) Si truncated microcones by tapered reactive ion etching (t-RIE). b) SEM image of Surface_10x15 after t-RIE. Scale bar 10 μm. c) Si truncated microcones covered by Al and immersed in hot water to obtain NanoAl. d) SEM image of NanoAl covering a microcone. Scale bar 500 nm.*

We cleaned the surfaces by dipping in acetone, isopropanol and deionized water, dried with a $N_2$ flow and activated the surfaces with oxygen plasma. Chemical vapour deposition (CVD) of 1H,1H,2H,2H-Perfluorodecyltriethoxysilane (Sigma-Aldrich) was performed by placing the samples and 200 μl of fluorosilane in a sealed (class IP-67) aluminium box (internal volume of 3.7 litres) heated at 150 °C for 3 h followed by an annealing for 1.5 h with the box opened (for covalently unbound silane removal).

The advancing and receding contact angles of microdroplets condensing and evaporating on the NanoAl were evaluated to be $\vartheta_a$=157 ±1° and $\vartheta_r$=145 ±6, respectively.[39]

*Experimental setup*

The condensation frosting experiments were carried out in a custom-built environmental chamber[39] and observed with a digital microscope (Dinolite AM7915MZTL - EDGE) from the top-view window. The field of view for observation of the condensation frosting process is 2724x2043 $\mu m^2$ and always centred in the middle of the patterned area, while for the final frost coverage we observed the entire sample. The single droplet self-ejection and coalescence induced self-ejection videos are recorded from the side view with a high speed camera (Phantom V640, Vision Research) coupled to a microscopy objective (50X Mitutoyo Plan Apo infinity corrected, long working distance=13 mm, resolving power=500 nm, depth of focus=900 nm) through a tube lens (InfiniTube Ultima). We illuminated the surfaces with a LED light (MULTILED QT, GSVITEC) placed outside the environmental chamber and on the back of the samples with respect to the high speed camera.

The plate temperature is set to $T_p$= -11 ° C. The sample and the temperature sensor (thin film PT100 thermocouple, RS pro, class B accuracy) are in thermal contact with the cold plate through a thermal pad (T-flex 600 Series Thermal Gap Filler, Laird Technologies, thickness of 1 mm, thermal conductivity of 3 W/mK). Given the low thermal inertia of the silicon samples (600 μm thick) and of the PT100, we assume the surface temperature $T_{surf} \cong T_p$. The saturation ratio is set to $s = P_{vap}/P_{vap,sat}(T_{surf})$= 2.8 where $P_{vap}$ is the water vapour pressure and $P_{vap,sat}(T_{surf})$ is the saturation vapour pressure at $T_{surf}$. All the frosting experiments are performed with samples positioned vertically so that jumping droplets that fall back to the surface are minimized. The procedure to assure the desired $T_{surf}$ from the beginning of the experiments is as follows: the samples are pressed from the edges on the thermal pad positioned on the plate at room temperature, the chamber closed and dry air is flushed until the vapor pressure inside the chamber falls below the saturation value corresponding to -11 °C; the $T_{surf}$ is lowered to -11 °C, humid air with a $P_{vap}$=7.4 hPa is fluxed and recording is started. Within a few seconds, the first condensation nuclei appear.

*Self-ejection percentage and L/d distributions*

We evaluated the effectiveness of all the surfaces in the self-ejection of individual condensation droplets of the first generation of nucleated droplets. For the videos with field of view of 2724x2043 $\mu m^2$ we extracted the frames and analysed 3÷4 square sub-zones with length of 500÷800 μm. By counting the number of self-ejected droplets and the total number of nucleated droplets we obtained the *percentage of self-ejecting droplets* ($\%_{s-ej}$). Every droplet nucleated on lateral walls or on the bottom self-ejects, while the ones on top of the truncated cones, which cannot self-eject, grow until they coalesce with other droplets on top or between the cones and jump off the surface. The Surface_5x10 was not evaluated for uncertainties related to the microscope resolution, but self-ejection was observed.

The droplet population of the most performant surface (10x20) was analysed in order to support the key idea of ice-bridging frustration by self-ejection. We used the ImageJ's plugin *Trainable Weka Segmentation*[44] to classify the images into "droplets" and "substrate" classes (Figure 2.a). The *Smooth*, *Threshold* and *Irregular watershed* plug-in were applied (Figure 2.b) and then *Analyze particles* to quantify the diameter and position of each droplet (Figure 2.c). The output was passed to a plugin realized for calculating the average size and distance between particles and their closest neighbors[45] that we modified to calculate $L/d$ for each drop, averaged over the 6 nearest neighbours. $L$ and $d$ are depicted in Figure 2.d, in accordance with the literature.[23] The normalized distribution of $L/d$ over the entire population is a sort of probability density function that a drop will be touched by ice bridges coming from its closest frozen neighbours. We also extracted the average distance (between the centres) of each droplet with its 6 neighbours and the diameter distributions.

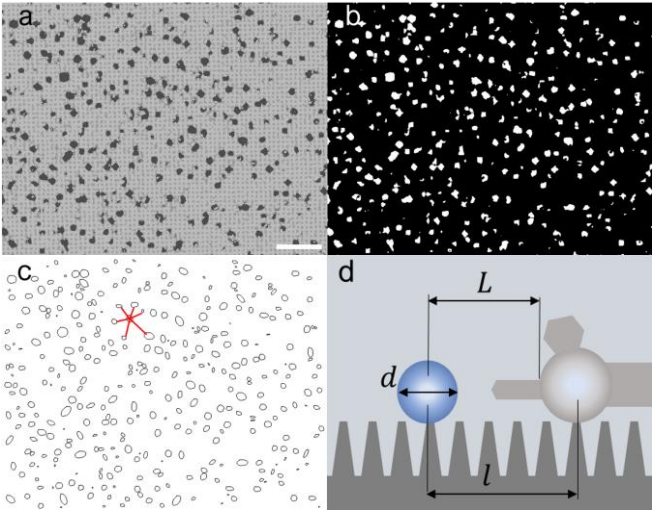

Figure 2. **a-c**) Sequence of the image analysis procedure performed with ImageJ of the Surface_10x20. Red lines in **c**) are an example of the calculations done by the code. The Analyse particles plugin fit each droplet with an ellipse. Each droplet radius is calculated as the mean of the two axes of the ellipse. **d**) Scheme of an ice bridge growing from a frozen droplet to a liquid one on top of the truncated microcones with $L$ and $d$ indicated.

## Frost velocity and coverage

In the literature there are several ways to measure the speed of frost propagation ($v_{frost}$) but there is not a univocal way. To calculate $v_{frost}$ on observation windows of few hundreds of microns, some authors divide either the length[24] or the diagonal[46] or the square root of the area[33] by the time in which all droplets in the observed area are frozen (frost velocity of the order of tens of µm/s). Other authors, quantify the frosting rapidity as an area covered by frost per second (µm²/s).[34,47] Others divide the area per second by the perimeter of the frost front (µm/s).[32,48] In our experiments, given the large field of view of the top microscope (2783 µm x 2087 µm), several fronts with different velocities are observed (non homogeneous spreading). The faster fronts follow paths along where $S$ is the smallest. In the other zones where $S \gg 1$ the droplets completely evaporate, an extended dry zone forms and the propagation velocity is negligible in comparison. For each experiment, we identified some (3÷6) directions along which the dendritic/ice-bridging front grows faster, measured their length ($l$) every 100 s, calculated their instantaneous velocity as $v = l/100s$ and the temporal mean $v_t$. We calculated $v_{frost}$ as the mean of the various $v_t$. We calculate $v_{frost}$ with only the faster fronts because they determines the millimetric scale connections between frost fronts started from different ice nucleation sites and are the weak link for the anti-frosting and anti-icing properties of a surface, e.g. the repellence to supercooled impacting droplets which requires a frost free large area. Indeed, as we will show, this $v_{frost}$ is specific of the population of droplets and not of the ice nucleation sites density ($\rho_{n,i}$) which can be affected by other parameters like impurities, surface defects and environmental conditions.[49] Important to note, this $v_{frost}$ is at best higher compared to the ones obtained with other methods and thus we can safely say that the present surfaces overcome the current limits.

We also calculated the global frost velocity $v_{glob}$ as the square root of the total patterned area divided by the time after which all the droplets are frozen, $t_{frost}$. We will argue about the relation between $v_{glob}$, $\rho_{n,i}$ and $v_{frost}$.

We used *Trainable Weka Segmentation* to classify the images into "frost" at the end of the propagation and "substrate" classes and then *Analyze Particles* to quantify the fraction of the area covered by frost. The *frosted area percentage* was evaluated for both the 2724x2043 µm² field of view in the centre of the sample ($A_{f,local}$) and at the scale of the entire sample ($A_{f,glob}$).

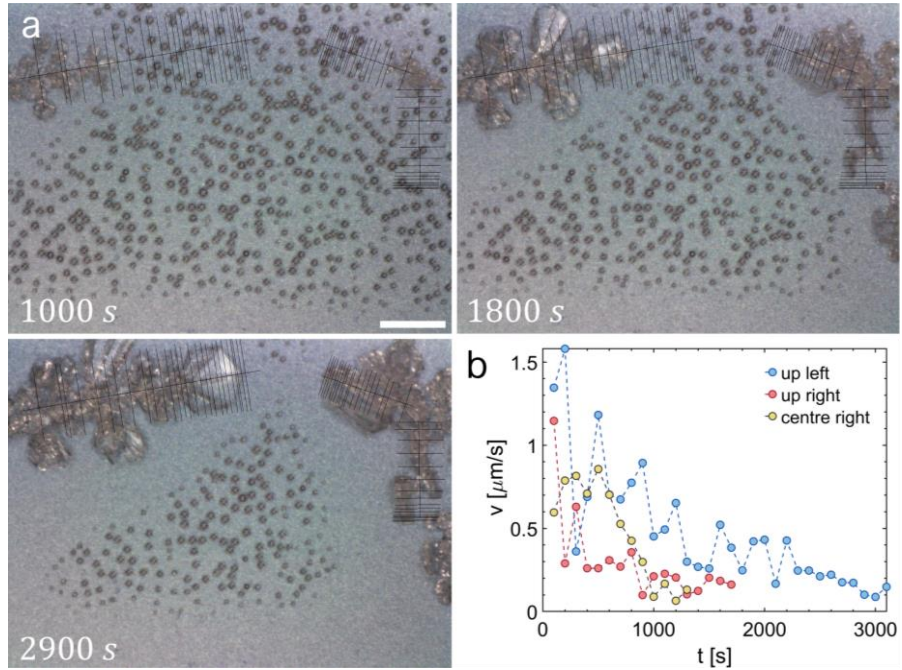

*Figure 3. **a**) Frost dendrites growth on Surface_5x10 with the black lines employed to measure $v$. On this surface, also at the beginning of frosting there is no ice-bridging but complete evaporation of the droplet with the formation of a large depletion zone. The scale bar is 400 µm. **b**) $v$ evolution for the three dendrites. It is higher at the beginning because the liquid droplets are closer to ice (the bridge velocity scales as $1/L$)[15] while when the dry zone widens it slows down. $v_{frost}$ = 0.42 ± 0.1 µm/s.*

## Results and Discussion

### *Self-ejection and $\%_{s-ej}$ vs $\varphi_H$*

As described analytically and proofed experimentally by the authors in a previous study, a droplet nucleated between the four microcones grows by condensation until it settles symmetrically. Then, the increasing volume accommodates to maintain a uniform internal pressure, thus a mechanical equilibrium, by variating the top and bottom menisci contact angles ($\vartheta_t$ and $\vartheta_b$) and heights ($H_t$ and $H_b$). Its shape is similar to a prolate spheroid and the correspondent surface energy ($E_s$) is not at its minimum. As the droplet reaches the dynamic configuration which corresponds to $\vartheta_t = \vartheta_a$, $\vartheta_b = \vartheta_r$ and the shape ratio $\lambda \equiv H_t/H_b = cos\,(\vartheta_a + \beta)/cos\,(\vartheta_r - \beta)$, it self-propels by releasing $E_s$. Its fast self-propulsion is determined by the driving surface force and the opposing capillary and viscous forces. The droplet accelerates, decelerates and stops at a certain height dependent on the dynamic angles and the tapering. Then, it continues to grow and the cycle repeats. Once it reaches the top edges of the truncated microcones, it grows to another dynamic configuration and self-ejects with a velocity dependent on the structures size, tapering and dynamic angles. The timeline is represented in Figure 4.a and a self-ejection event in Figure 4.b (see Supplementary video 1).

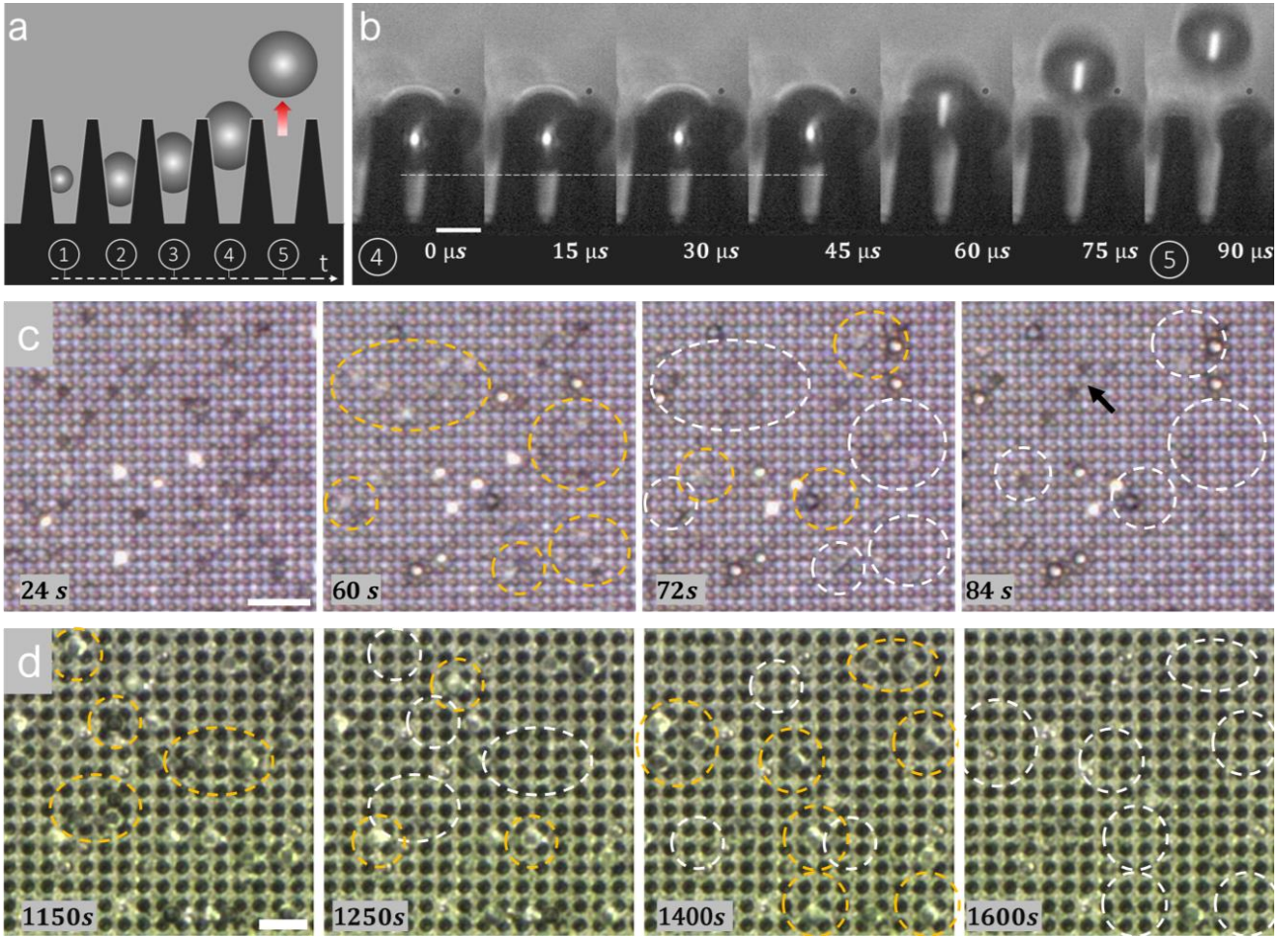

*Figure 4. **a**) The timeline of a condensation droplet from nucleation to self-ejection. The droplet 1) nucleates, 2) settles between the four cones and grows to the dynamic configuration, 3) alternates fast self-propulsions to slow growths until it reaches the top of the truncated cones and 4) grows to a self-ejection dynamic configuration. From 4) to 5) it self-ejects as captured in **b**) in tens or hundreds of microseconds depending on the structures' size, while from phase 1) to 4) it can take tens of seconds to minutes. The captured droplet self-ejects from Surface_10x13. Scale bar: 10 µm. **c**) Top view experiment of the Surface_15x20p. The dotted yellow ellipses indicate where self-ejection is going to occur and the white ones in the next frame where droplets self-ejected. The dark zones indicated by the black arrow at 84 s are re-nucleation sites. The time displayed refers to the beginning of the video which coincides with the introduction of humid air. **d**) Self-ejection events from Surface_30x40.*

In Figure 4.c-d we report two examples of top-view recordings (see Supplementary video 2-3). Every droplet nucleated on the side-walls or bottoms of the truncated cones self-ejects, while the ones on the heads of the cones can only jump off due to coalescence. Figure 5 reports the $\%_{s-ej}$ as $\varphi_H$ varies for all the surfaces. The trend confirms this description: the higher $\varphi_H$, the fewer the self-ejections. Thanks to self-ejection, the few droplets on the heads of the microcones are very spaced and rarely as large to have $S^* \lesssim 1$ also because they can jump by coalescing with other droplets (which will be discussed in the next paragraph). Large droplets spacing and small diameters implies $S^* > 1$ for every droplet and thus frustration of ice bridges.

The slightly different nanostructuring procedure between Group1 and Group2 has no effect on enabling self-ejection and thus on the relations with $\varphi_H$ analysed in the present work; every droplet not nucleated on the heads, self-ejects.

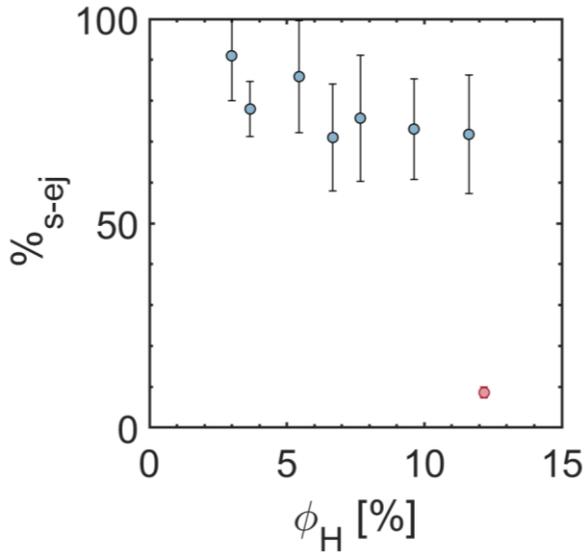

*Figure 5. Percentage of self-ejecting droplets (%$_{s-ej}$) with respect to the total number of droplets nucleated as the head area percentage, $\varphi_H$, varies.*

## *Other jumping modes*
After self-ejection there are two possibilities:
1) droplets re-nucleate on the same site, grow and self-eject; the cycle repeats.
2) the old nucleation site remains free and condensation accumulates on neighbouring droplets placed between or on top of the cones.

After a certain time that depends on cone spacing and therefore on the time to achieve the self-ejection volume, the remaining droplets are only those nucleated on the heads of the cones. They can leave the surface too and in the case of a vertical surface there are 3 ways: they coalesce either i) with other drops on top of the heads or ii) with those located between the cones that grow and self-propel towards the aperture (see Figure 6.a, an event captured in slow motion), or iii) because they are swept away by droplets located between the cones in adjacent unit cells that coalesce and eject (Figure 6.b). For completeness, we performed some experiments with the surfaces placed horizontally: droplets that fall back after self-ejection or coalescence jumping will trigger other cascading ejections induced by coalescence (Figure 6.c). The events of Figure 6.a-c are reported in the Supplementary videos 4-6, respectively.

Surface_60x80 (with the highest $\varphi_H$, reported in Figure 6.c) is an exception compared to the other surfaces because frost freezes the drops between the cones before they can reach the self-ejection volume; however, there were coalescence jumps (of the various types presented above) which in this case are more frequent than on the other surfaces and begin before self-ejection. To avoid frost and give a value to the %$_{s-ej}$ we tested the surface with s = 2.8 but at 1 °C: it corresponds to the yellow point in Figure 5, %$_{s-ej}$ is less than 10% and coalescence jumping is dominant. This is due to a number of condensate nuclei per unit area greater than the number of cells of four cones per unit area. Thus, to maximise self-ejection for a large spectrum of saturation ratio one should design the smaller microstructures possible, compatibly with self-ejection requirements.

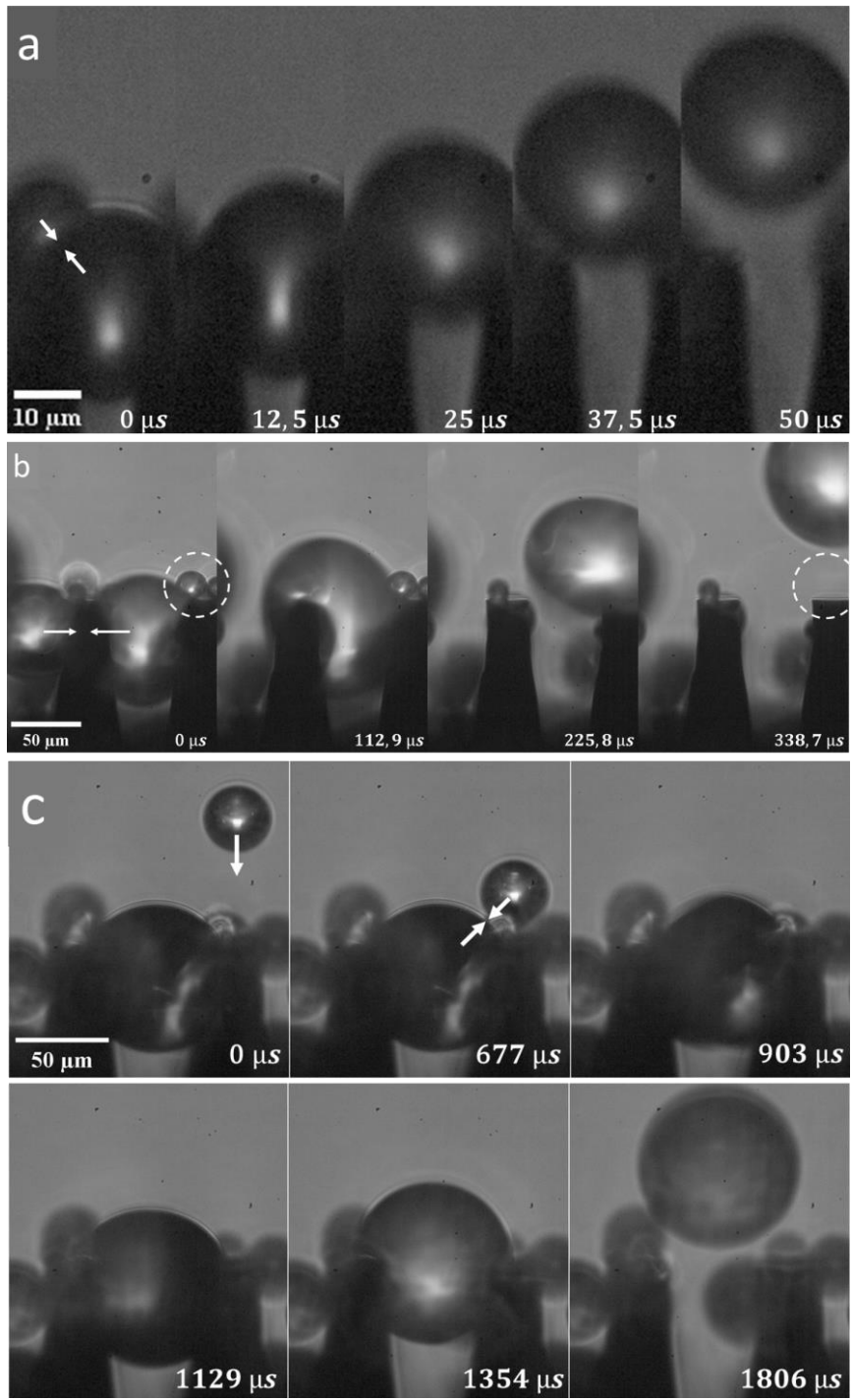

*Figure 6. **a**) Ejection of a droplet located between the microcones induced by coalescence with a droplet on top (Surface_15x20). **b**) Coalescence and ejection of two droplets located between the microcones that also swept away two droplets on top of the right cone (Surface_60x80). **a**) and **b**) are typical jumping modes alternative to self-ejection that happen on vertical surfaces. **c**) Additional jumping mode on Surface_60x80 placed horizontally. Droplets that jumped (due to self-ejection or coalescence) can fall back on the surface and trigger cascading ejections induced by coalescence.*

### Frost velocity

The most performant surfaces (Figure 3.a and 7.a) correspond to the higher $\%_{s-ej}$; as soon as a drop spontaneously freezes in the field of view or the frost front arrives from outside of it, all the drops evaporate completely and the frost advances with dendritic formations by desublimation. The crystals are strongly anisotropic because certain facets grow faster than others.

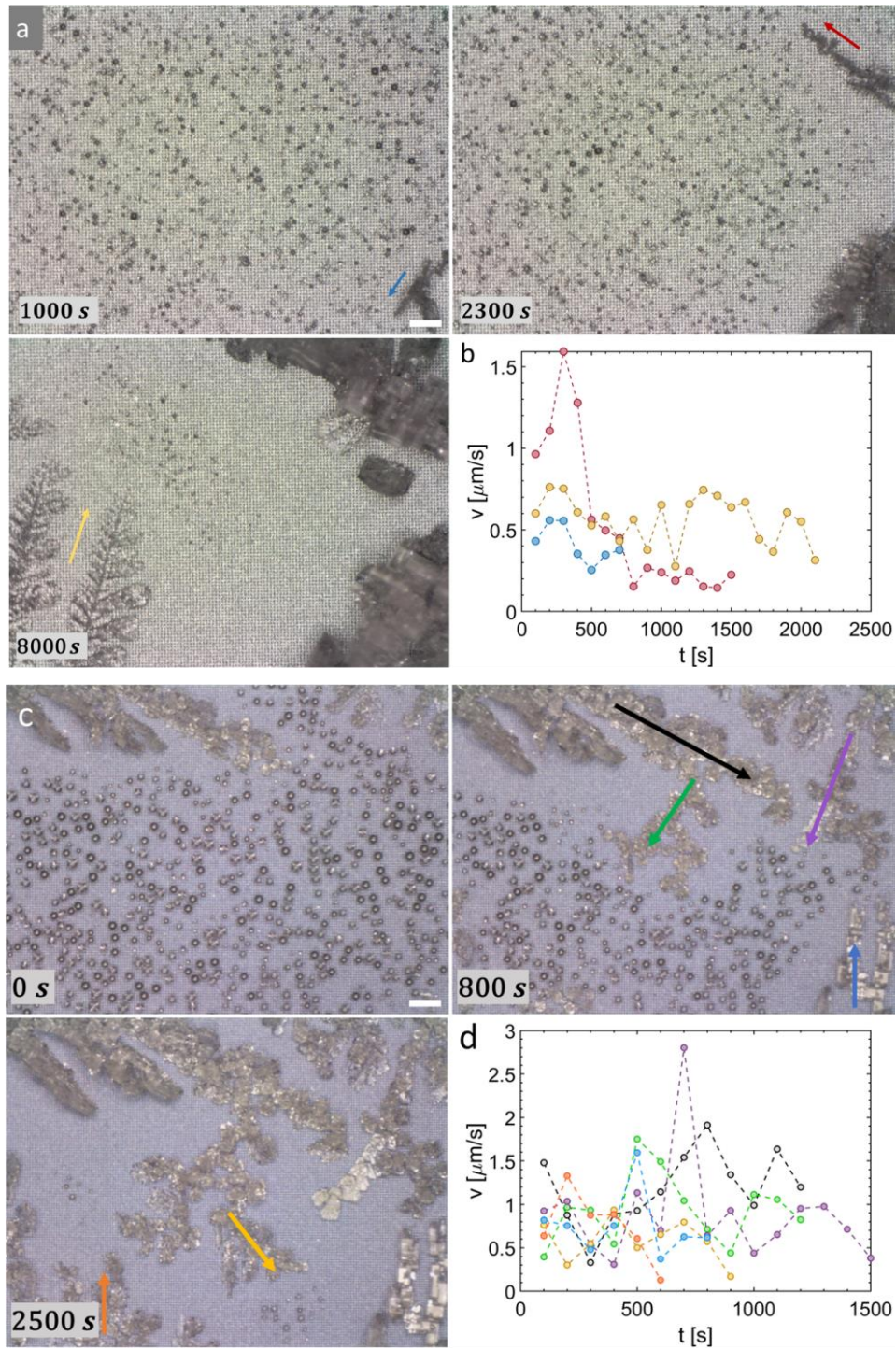

*Figure 7. **a**) Dendrites growth on Surface_10x20. Ice-bridging is suppressed for all of them. The fastest fronts are identified by coloured arrows. **b**) $v$ with the correspondent colour. We plotted $v$ with the same starting time even if they grow in different time periods. $v_{frost}$= 0.50 ± 0.08 μm/s. **c**) Surface_10x15 with coloured arrows correspondent to $v$ in **d**). The frost grows with a mixed mode. $v_{frost}$= 0.85 ± 0.21 μm/s. Scale bar 200 μm.*

A clear example is in Figure 7.a (see Supplementary video 6): the two initial ice crystals (blue and red arrows) have an elongated shape; the short side grows more "rapidly" towards the drops while the long one is much slower (about one third) and forms a large dry zone. The slow facets are not considered in the calculation of $v$. Moreover, the fast side has a typical slow down of $v$ as the dry zone develops (as also in Figure 3). The two long dendrites developed after about 5000 s (yellow arrow in Figure 7.a) are instead more branched, faster and maintain their speed. In fact, they determine the complete evaporation of the remaining condensation. However, also in their case successful ice-bridging was not observed. Therefore the $v$ evolution depends also on the crystal shape (Figure 7.b). As shown in Kenneth G. Libbrecht's *Snow Crystals*[50], the shape and growth rate of ice crystals depends on various parameters such as temperature and saturation ratio. As the temperature is constant during the experiments, we deduce that the saturation ratio is altered in the yellow arrow zone by the presence of large dry zones and frost changes the distribution of vapour fluxes from the surrounding air. On other surfaces with larger $\varphi_H$ and smaller $\%_{s-ej}$, we instead observed a mixed behaviour, i.e. both ice-bridging and the formation of dry zones (for example in Figure 7.c). This can be seen from the $v$ of various paths in Figure 7.d: unlike the cases with full ice-bridge frustration where velocity steadily decreases, in the mixed behaviour it oscillates.

The self-ejection phenomenon has an anti-frosting effect superior than coalescence jumping alone because it is independent on the neighbour droplets. The illustrations in Figure 8.a-b depict the condensation frosting on surfaces that enable self-ejection: the majority of droplets leave the surface as soon as they reach the self-ejection volume and the remaining ones, on the microcones' heads, are far from each other and evaporate completely due to the presence of ice crystals. Thus, the ice-bridging propagation mechanism is frustrated.

Let's suppose a certain number of condensation nuclei per unit area, with density $\rho_c$ higher than the four-microcones unit cell density, so that for each cell there is a maximum of one droplet. Assuming a population of equally spaced drops, let's estimate the mean distance between the droplets centres as $l_c \sim \sqrt{1/\rho_c}$. In the limit case of ideal microcones, where $\varphi_H \to 0$ and $\%_{s-ej} \to 100\%$, each condensation droplet self-ejects at a precise diameter $D_{ej}$. This diameter is dependent on the size, tapering and dynamic angles of the microstructures,[39] which also define the maximum droplet diameter. Considering the dimensions depicted in Figure 2.d, re-adapted to the case of droplets between the cones and in the most favourable scenario for successful ice-bridging (at the time of maximum diameter), $L \sim l_c - D_{ej}/2$. By imposing $S^* \equiv L/D_{ej} > 1$, we obtain the criterion $D_{ej} \lesssim 2l_c/3$. Therefore, for a certain $\rho_c$, dependent on the environmental conditions and the surface roughness and chemistry, there is a critical self-ejection diameter $D_{ej}^* = 2l_c/3$, tuneable in the design process, under which all the ice bridges fail.

Now let's consider truncated microcones with a certain $\%_{s-ej} < 100\%$; we can consider the droplets in two classes: 1) the self-ejecting ones that assure the ice bridges frustration if the structures are properly designed ($D_{ej} < D_{ej}^*$) and 2) the ones on the heads that can potentially let the frost propagate by contact before they coalesce and jump off. The first generation droplets remaining on the heads have a density $\rho_{c,H} = \rho_c(1 - \%_{s-ej}/100)$ and we can say from the experiments (last paragraph) that it is steady because of the various jumping interactions with the droplets of subsequent generations. The mean distance between their centres becomes $l_{c,H} \sim \sqrt{1/\rho_{c,H}}$. Thus, the larger the $\%_{s-ej}$ (or, in other words, the smaller the $\varphi_H$), the larger the $l_{c,H}$ and the mean $S^*$. In the last section we shall show that $S^* > 1$ for all the droplets on one of the most performant surfaces.

By plotting $v_{frost}$ as $\varphi_H$ varies (Figure 8.c), we see again an approximately linear dependence which is implicitly affected by the $\%_{s-ej}$, as explained. For surfaces with the highest $\%_{s-ej}$ we recorded the lowest frost velocities. $v_{glob}$ is at least $2v_{frost}$ as there are multiple fronts advancing in opposite directions starting from the edges of the patterned sample, where there are defects and ice forms quickly because of the hydrophilic plate, and from spontaneous ice nuclei. On a ideally infinite array of nanostructured truncated microcones (to neglect sample edges), with the presence of ice nuclei distributed with a mean distance $l_{n,i} \sim \sqrt{1/\rho_{n,i}}$, one can estimate the frosting time as $t_{frost} \gtrsim l_{n,i}/2v_{frost}$, where the equal stands for the (rare) simultaneous appearance of the nuclei. In the presence of icephilic edges of a sample of size $l_s$ and no other ice nuclei, $t_{frost} \sim l_s/2v_{frost}$. In summary, being $v_{frost}$ a quantity referred to the propagation from one ice nucleation site towards a population of jumping droplets (self-ejection and coalescence jumping) which extends for several hundreds of microns, it is representative to compare it with most of the cited studies that report the frost velocity relative to areas with a side length of a few hundred microns. We can thus say that surfaces that enable self-ejection cause frost to propagate 2 to 6 times slower than the state of art surfaces.[24,33] $v_{glob}$ is instead a quantity dependent on both $v_{frost}$ and $l_{n,i}$.

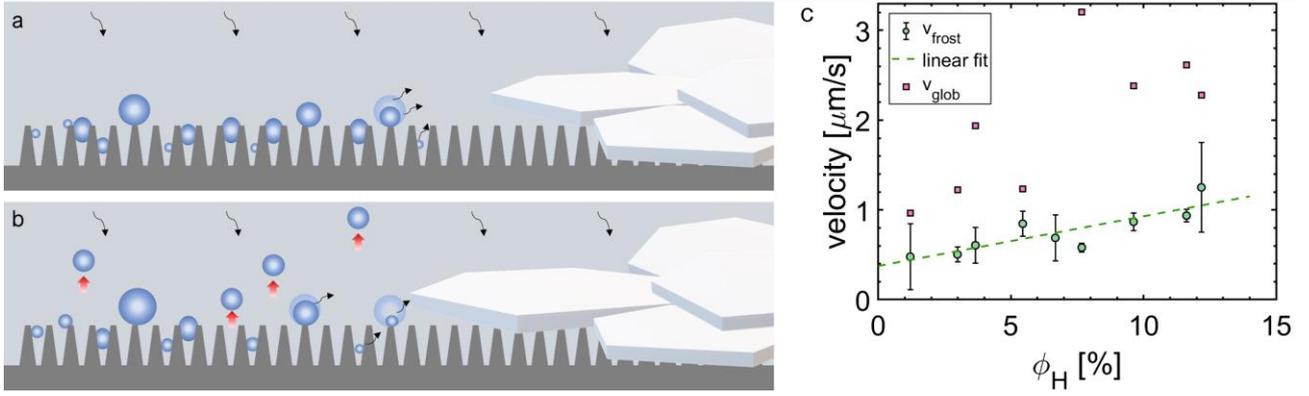

*Figure 8. **a**-**b**) Illustration of two subsequent times during condensation frosting on nanostructured truncated microcones that enable self-ejection. An high percentage of droplets leave the surfaces at a precise volume and the remaining ones are far enough to evaporate completely before being touched by ice bridges or crystals. **c**) $v_{frost}$ and $v_{glob}$ as $\varphi_H$ varies. The linear fit is $v_{frost} = 0.0554 \cdot \varphi_H + 0.374$ and has a $R^2$=0.73. $v_{glob} \sim 2 v_{frost}$ when frost starts form the edges and higher when there are multiple ice nuclei.*

## *Frost coverage*

Regarding the surface's frost coverage we still found an improvement compared to surfaces that exhibit pure ice-bridging or a frost delay only due to coalescence jumping[24,30–36]. As seen in Figure 9, $A_{f,local}$ goes from 20 to 50% while $A_{f,glob}$ goes from 41 to 68%; On the scale of the entire sample the edges have an effect because it can happen that frost starts from there and reaches the drops by ice-bridging before they leave the surface by self-ejection. On larger surfaces we expect $A_{f,glob} \to A_{f,local}$. In any case, the free final area is high compared to ice-bridging surfaces or those that exhibit coalescence jumping. Looking for applications, surfaces that exploit self-ejection could improve the efficiency of the heat exchangers where the frost coverage is a determining factor. For airplanes, a part from delaying the frosting, a surface that enables self-ejection would have a large area free from frost, still superhydrophobic; supercooled impacting droplets could thus rebound on the free area and wet only frost and freeze. The delamination of the formed ice layer (by gravity, wind and other means) would be facilitated because it would have a contact area fraction of $f * A_{f,local}/100$, with $f$ a coefficient defined as the ratio between the real solid-ice area and $A_{f,local}$ which is a projected area.

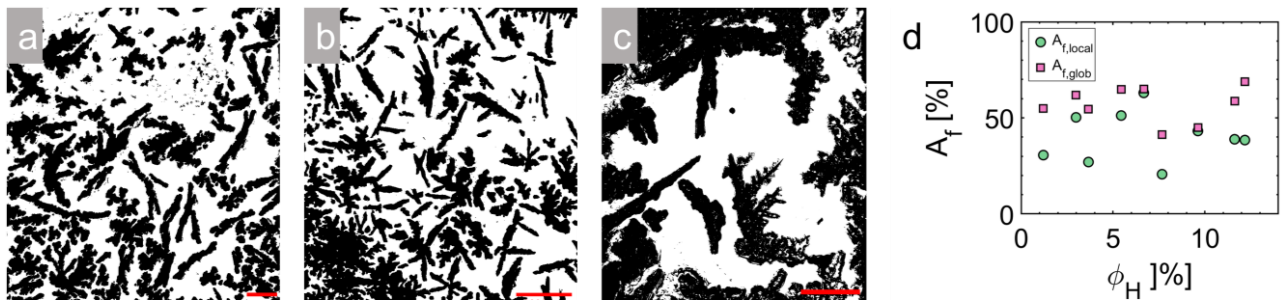

*Figure 9. From **a**) to **c**), the frost on the patterned area (global) of Surface_15x20, Surface_30x40, and Surface_10x20, respectively. **d**) $A_{f,glob}$ and $A_{f,local}$ as $\varphi_H$ varies. Scale bars 2 mm.*

*Ice-bridging frustration by self-ejection*

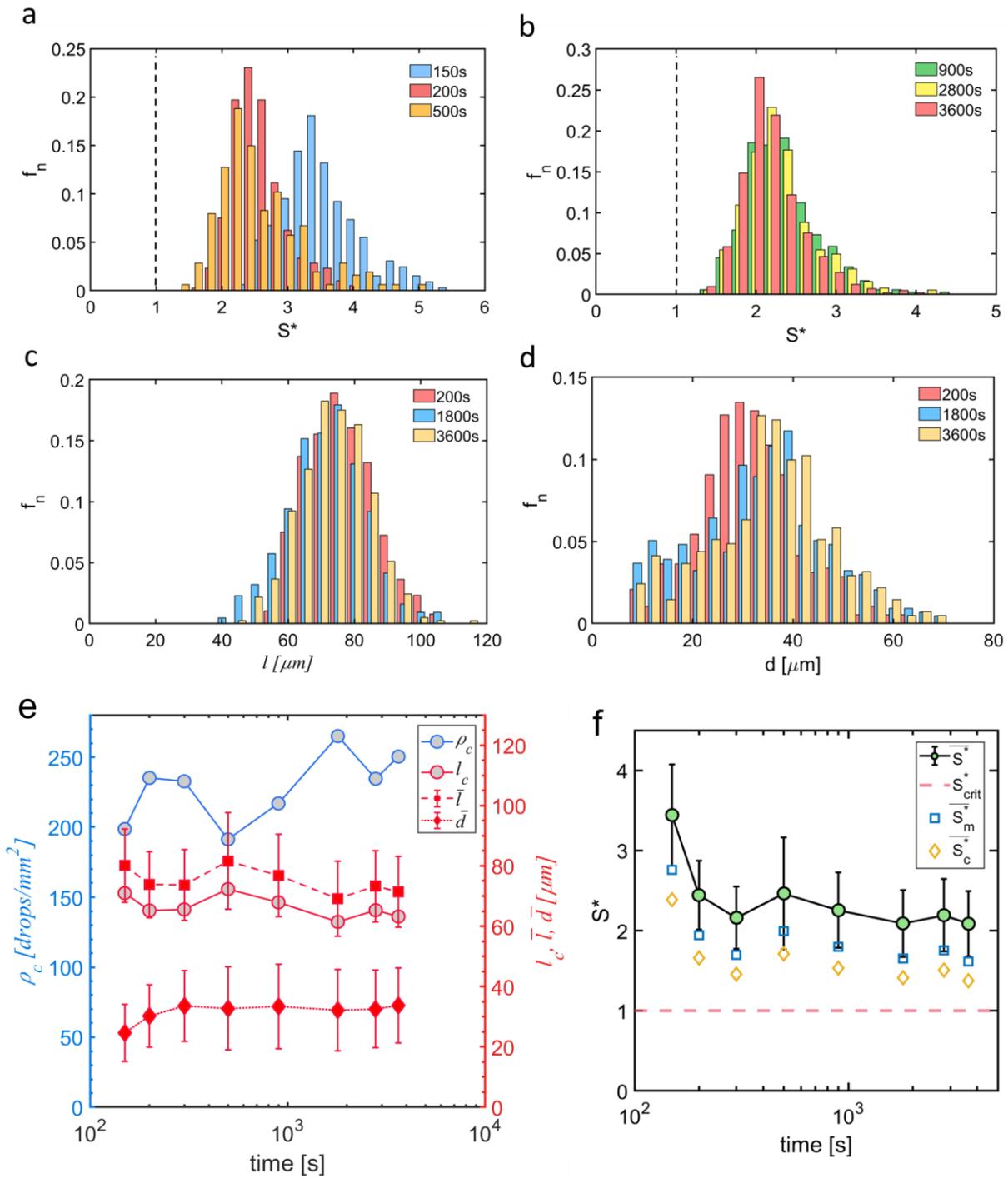

Figure 10. Normalized frequency $f_n$ of **a-b)** the bridging parameter $S^*$ (bin size 0.2) and of **c)** the centres distance $l$ (bin size 5 μm) calculated as the average among the six nearest neighbours of each droplet. **d)** Normalized frequency $f_n$ of the droplet diameter $d$ (bin size 3 μm). The times are referred to the beginning of the experiment. **e)** $\rho_c$, $l_c$ (calculated as $\sqrt{1/\rho_c}$) and the mean and standard deviation of the $l$ and $d$ distributions. **f)** Mean, $\overline{S^*}$, and standard deviation of the $S^*$ distributions. There are no droplets that satisfy the condition $S^* < S^*_{crit}=1$ and indeed the ice-bridging in frustrated. The estimates of $\overline{S^*}$, $\overline{S^*}_m$ and $\overline{S^*}_c$, are plotted for comparison. All the data are obtained by analysing the droplets on Surface_10x20 in a field of view of area 1280x12800 μm².

We verified that the frustration of ice bridging for all the droplets is supported by the condition $S^* > 1$ for all the droplets population. In Figure 10.a-b we show the normalized distributions of $S^*$ of

Surface_10x20 at six times in the range 150s÷3600s after the beginning of the experiment and before any drop freezes in the field of view. After a few minutes, the drops begin to jump for self-ejection in a recurrent way from the same sites, also causing other kinds of jumping. The continuous self-ejections and coalescence jumps between droplets located between and on top of the cones cause the distributions to be stationary and with zero frequency for $S^*<1$. The jumping mode of coalescence of two/more droplets on the head is rare because the coalescence with the ones between the cones is highly likely. Also the distributions of distances between the droplet centres ($l$) and droplet diameters ($d$) are steady (Figure 10.c-d). The number of drops per unit area (Figure 10.e) becomes quite stationary after ~10 mins and oscillates in the range 210÷260 drops/mm². We see that $l_c$, calculated as $\sqrt{1/\rho_c}$, gives an acceptable estimate of the distance of the droplet centres averaged among the six nearest neighbours for each droplet, $l$. In particular, the trend over time is the same and $l/l_c \sim 1.12$. The mean diameter is also steady. Finally, Figure 10.f shows a steady mean of $S^*$, $\overline{S^*}$, relative to the distributions in Figure 10.a-b. However, the approximate parameters that one can calculate as $\overline{S^*}_m \approx (\bar{l} - \bar{d}/2)/\bar{d}$ or $\overline{S^*}_c \approx (l_c - \bar{d}/2)/\bar{d}$, underestimate $\overline{S^*}$ by a relative error of -21% and -32%, respectively. Therefore, although the time trend is consistent, it is preferable to compute $\overline{S^*}$ from its real distribution.

From the point of view of surfaces that improve heat exchange by condensation, where small droplets are preferred, the distribution of droplet diameters (Figure 10.d) is worthy of note, as it can be strictly controlled by the self-ejection diameter (~28 μm for Surface_10x20) which is a designable parameter.

# Perspective and conclusion

We have shown that single droplet self-ejection delays the frost propagation and overcomes the performance of the state of the art surfaces of up to one order of magnitude. On the most performing surfaces where self-ejection is predominant the frost propagation by successful ice bridges is completely frustrated ($S^*>1$ for all the droplets) and the frost propagates with velocities in the range 0.4÷1.2 $\mu m/s$, from 2 to 6 times slower the state of art surfaces. The formation of large dry areas by evaporation of the drops brings the final frost coverage in the range 41÷68%, depending on the location and scale considered.

We also related both the percentage of self-ejection and the $v_{frost}$ to the cone heads area percentage $\varphi_H$. The smaller the $\varphi_H$, the larger the %$_{s-ej}$ and the smaller the $v_{frost}$. Ideal cones ($\varphi_H \rightarrow 0$) are expected to provide a %$_{s-ej}$=100% and reduce $v_{frost}$ even more. Future studies will go in this direction.

Regarding the materials used, manufacturing on silicon with precise tuning of size and tapering served to prove the concept of self-ejection and its anti-frosting effect. Future works will be oriented on the realization of conical microstructures on materials commonly used for heat exchangers such as aluminium which already offers the great advantage of an easy and scalable nanostructuring or on soft polymers, intrinsically hydrophobic, which would eliminate the fragile silane layer.

Regarding the structure geometry, the tapered shape promises greater advantages compared to the vertical one (pillars or grooves) because it can provide a certain mechanical resistance even if $\varphi_H \rightarrow 0$ which we proofed to be optimal for %$_{s-ej}$ and $v_{frost}$. Vertical structures have instead an intrinsic limit on their thickness, thus on $\varphi_H$, due to mechanical stability restrictions. Optimal $\varphi_H$ and $\beta$ will be investigated.

Future studies will document and analyse the effect of different dynamic angles and of tapering on self-ejection velocity up to the limits after which self-ejection is inhibited.

In conclusion, self-ejection promises enhancements also in other fields of application where coalescence jumping has already provided improvements, such as condensation heat transfer, water harvesting and self-cleaning among others.

## Acknowledgements

We kindly thanks Alberto Bellin, Luigi Fraccarollo and Fabio Sartori (University of Trento, Italy) who lent us the Phantom camera and the Micro Nano Facility technical staff (FBK, Trento, Italy) for support in the fabrication. The study has been supported by the European Commission under the FET Open "Boheme" grant no. 863179.

## Author contributions

N.G.D. and A.B designed the study and fabricated the surfaces. N.G.D. tested the surfaces, interpreted the data and wrote the manuscript. A.B. and N.M.P. discussed the data and edited the manuscript. A.B. and N.M.P. supervised the research.

## Supplementary information

See the Supplementary information document

## References

(1) Rafati Nasr, M.; Fauchoux, M.; Besant, R. W.; Simonson, C. J. A Review of Frosting in Air-to-Air Energy Exchangers. *Renew. Sustain. Energy Rev.* **2014**, *30*, 538–554. https://doi.org/10.1016/j.rser.2013.10.038.

(2) da Silva, D. L.; Melo, C.; Hermes, C. J. L. Effect of Frost Morphology on the Thermal-Hydraulic Performance of Fan-Supplied Tube-Fin Evaporators. *Appl. Therm. Eng.* **2017**, *111*, 1060–1068. https://doi.org/10.1016/j.applthermaleng.2016.09.165.

(3) Xia, Y.; Zhong, Y.; Hrnjak, P. S.; Jacobi, A. M. Frost, Defrost, and Refrost and Its Impact on the Air-Side Thermal-Hydraulic Performance of Louvered-Fin, Flat-Tube Heat Exchangers. *Int. J. Refrig.* **2006**, *29* (7), 1066–1079. https://doi.org/10.1016/j.ijrefrig.2006.03.005.

(4) Lynch, F. T.; Khodadoust, A. Effects of Ice Accretions on Aircraft Aerodynamics. *Prog. Aerosp. Sci.* **2001**, *37* (8), 669–767. https://doi.org/10.1016/S0376-0421(01)00018-5.

(5) Mahvi, A. J.; Boyina, K.; Musser, A.; Elbel, S.; Miljkovic, N. Superhydrophobic Heat Exchangers Delay Frost Formation and Enhance Efficency of Electric Vehicle Heat Pumps. *Int. J. Heat Mass Transf.* **2021**, *172*. https://doi.org/10.1016/j.ijheatmasstransfer.2021.121162.

(6) Li, K.; Xia, D.; Luo, S.; Zhao, Y.; Tu, R.; Zhou, X.; Zhang, H.; Su, L. An Experimental Investigation on the Frosting and Defrosting Process of an Outdoor Heat Exchanger in an Air Conditioning Heat Pump System for Electric Vehicles. *Appl. Therm. Eng.* **2022**, *201* (June 2021). https://doi.org/10.1016/j.applthermaleng.2021.117766.

(7) Bragg, M. B.; Broeren, A. P.; Blumenthal, L. A. Iced-Airfoil Aerodynamics. *Prog. Aerosp. Sci.* **2005**, *41* (5), 323–362. https://doi.org/10.1016/j.paerosci.2005.07.001.

(8) Kent, R. A.; Andersen, D.; Caux, P. Y.; Teed, S. Canadian Water Quality Guidelines for Glycols - An Ecotoxicological Review of Glycols and Associated Aircraft Anti-Icing and Deicing Fluids. *Environ. Toxicol.* **1999**, *14* (5), 481–522. https://doi.org/10.1002/(SICI)1522-7278(199912)14:5<481::AID-TOX5>3.0.CO;2-8.

(9) Goraj, Z. An Overview of the Deicing and Antiicing Technologies with Prospects for the Future. *24Th Int. Congr. Aeronaut. Sci.* **2004**, 1–11.

(10) Su, Q.; Chang, S.; Zhao, Y.; Zheng, H.; Dang, C. A Review of Loop Heat Pipes for Aircraft Anti-Icing Applications. *Appl. Therm. Eng.* **2018**, *130*, 528–540. https://doi.org/10.1016/j.applthermaleng.2017.11.030.

(11) Oberli, L.; Caruso, D.; Hall, C.; Fabretto, M.; Murphy, P. J.; Evans, D. Condensation and Freezing of Droplets on Superhydrophobic Surfaces. *Adv. Colloid Interface Sci.* **2014**, *210*, 47–57. https://doi.org/10.1016/j.cis.2013.10.018.

(12) Maitra, T.; Tiwari, M. K.; Antonini, C.; Schoch, P.; Jung, S.; Eberle, P.; Poulikakos, D. Erratum: Nanoengineering of Superhydrophobic and Impalement Resistant Surface Textures below the Freezing Temperature (Nano Letters (2014) 14:1 (172-182) DOI: 10.1021/Nl4037092). *Nano Lett.* **2014**, *14* (2), 1106. https://doi.org/10.1021/nl500297b.

(13) Zuo, Z.; Song, X.; Liao, R.; Zhao, X.; Yuan, Y. Understanding the Anti-Icing Property of Nanostructured Superhydrophobic Aluminum Surface during Glaze Ice Accretion. *Int. J. Heat Mass Transf.* **2019**, *133*, 119–128. https://doi.org/10.1016/j.ijheatmasstransfer.2018.12.092.

(14) Eberle, P.; Tiwari, M. K.; Maitra, T.; Poulikakos, D. Rational Nanostructuring of Surfaces for Extraordinary Icephobicity. *Nanoscale* **2014**, *6* (9), 4874–4881. https://doi.org/10.1039/c3nr06644d.

(15) Nath, S.; Boreyko, J. B. On Localized Vapor Pressure Gradients Governing Condensation and Frost Phenomena. *Langmuir* **2016**, *32* (33), 8350–8365. https://doi.org/10.1021/acs.langmuir.6b01488.

(16) Jo, H.; Hwang, K. W.; Kim, D.; Kiyofumi, M.; Park, H. S.; Kim, M. H.; Ahn, H. S. Loss of Superhydrophobicity of Hydrophobic Micro/Nano Structures during Condensation. *Sci. Rep.* **2015**, *5*, 5–10. https://doi.org/10.1038/srep09901.

(17) Ryan Enright,, Nenad Miljkovic, Ahmed Al-Obeidi, Carl V. Thompson, and E. N. W. Condensation on Superhydrophobic

Surfaces: The Role of Local Energy Barriers and Structure Length Scale. *Langmuir* **2012**, *28*.

(18) Guadarrama-Cetina, J.; Mongruel, A.; González-Viñas, W.; Beysens, D. Percolation-Induced Frost Formation. *Epl* **2013**, *101* (1). https://doi.org/10.1209/0295-5075/101/16009.

(19) Chen, X.; Ma, R.; Zhou, H.; Zhou, X.; Che, L.; Yao, S.; Wang, Z. Activating the Microscale Edge Effect in a Hierarchical Surface for Frosting Suppression and Defrosting Promotion. *Sci. Rep.* **2013**, *3*, 1–8. https://doi.org/10.1038/srep02515.

(20) Varanasi, K. K.; Deng, T.; Smith, J. D.; Hsu, M.; Bhate, N. Frost Formation and Ice Adhesion on Superhydrophobic Surfaces. *Appl. Phys. Lett.* **2010**, *97* (23). https://doi.org/10.1063/1.3524513.

(21) Schutzius, T. M.; Jung, S.; Maitra, T.; Eberle, P.; Antonini, C.; Stamatopoulos, C.; Poulikakos, D. Physics of Icing and Rational Design of Surfaces with Extraordinary Icephobicity. *Langmuir* **2015**, *31* (17), 4807–4821. https://doi.org/10.1021/la502586a.

(22) Chavan, S.; Park, D.; Singla, N.; Sokalski, P.; Boyina, K.; Miljkovic, N. Effect of Latent Heat Released by Freezing Droplets during Frost Wave Propagation. *Langmuir* **2018**, *34* (22), 6636–6644. https://doi.org/10.1021/acs.langmuir.8b00916.

(23) Nath, S.; Ahmadi, S. F.; Boreyko, J. B. How Ice Bridges the Gap. *Soft Matter* **2020**, *16* (5), 1156–1161. https://doi.org/10.1039/c9sm01968e.

(24) Boreyko, J. B.; Collier, C. P. Delayed Frost Growth on Jumping-Drop. *ACS Nano* **2013**, No. 2, 1618–1627.

(25) Lv, C.; Hao, P.; Yao, Z.; Song, Y.; Zhang, X.; He, F. Condensation and Jumping Relay of Droplets on Lotus Leaf. *Appl. Phys. Lett.* **2013**, *103* (2). https://doi.org/10.1063/1.4812976.

(26) Wang, F. C.; Yang, F.; Zhao, Y. P. Size Effect on the Coalescence-Induced Self-Propelled Droplet. *Appl. Phys. Lett.* **2011**, *98* (5). https://doi.org/10.1063/1.3553782.

(27) Mulroe, M. D.; Srijanto, B. R.; Ahmadi, S. F.; Collier, C. P.; Boreyko, J. B. Tuning Superhydrophobic Nanostructures to Enhance Jumping-Droplet Condensation. *ACS Nano* **2017**, *11* (8), 8499–8510. https://doi.org/10.1021/acsnano.7b04481.

(28) Lecointre, P.; Mouterde, T.; Checco, A.; Black, C. T.; Rahman, A.; Clanet, C.; Quéré, D. Ballistics of Self-Jumping Microdroplets. *Phys. Rev. Fluids* **2019**, *4* (1), 24–26. https://doi.org/10.1103/PhysRevFluids.4.013601.

(29) Wasserfall, J.; Figueiredo, P.; Kneer, R.; Rohlfs, W.; Pischke, P. Coalescence-Induced Droplet Jumping on Superhydrophobic Surfaces: Effects of Droplet Mismatch. *Phys. Rev. Fluids* **2017**, *2* (12), 1–17. https://doi.org/10.1103/PhysRevFluids.2.123601.

(30) Shen, Y.; Jin, M.; Wu, X.; Tao, J.; Luo, X.; Chen, H.; Lu, Y.; Xie, Y. Understanding the Frosting and Defrosting Mechanism on the Superhydrophobic Surfaces with Hierarchical Structures for Enhancing Anti-Frosting Performance. *Appl. Therm. Eng.* **2019**, *156* (February), 111–118. https://doi.org/10.1016/j.applthermaleng.2019.04.052.

(31) Zhao, G.; Zou, G.; Wang, W.; Geng, R.; Yan, X.; He, Z.; Liu, L.; Zhou, X.; Lv, J.; Wang, J. Rationally Designed Surface Microstructural Features for Enhanced Droplet Jumping and Anti-Frosting Performance. *Soft Matter* **2020**, *16* (18), 4462–4476. https://doi.org/10.1039/d0sm00436g.

(32) Chu, F.; Lin, Y.; Yan, X.; Wu, X. Quantitative Relations between Droplet Jumping and Anti-Frosting Effect on Superhydrophobic Surfaces. *Energy Build.* **2020**, *225*, 110315. https://doi.org/10.1016/j.enbuild.2020.110315.

(33) Li, L.; Lin, Y.; Rabbi, K. F.; Ma, J.; Chen, Z.; Patel, A.; Su, W.; Ma, X.; Boyina, K.; Sett, S.; Mondal, D.; Tomohiro, N.; Hirokazu, F.; Miljkovic, N. Fabrication Optimization of Ultra-Scalable Nanostructured Aluminum-Alloy Surfaces. *ACS Appl. Mater. Interfaces* **2021**. https://doi.org/10.1021/acsami.1c08051.

(34) Kim, A.; Lee, C.; Kim, H.; Kim, J. Simple Approach to Superhydrophobic Nanostructured Al for Practical Antifrosting Application Based on Enhanced Self-Propelled Jumping Droplets. *ACS Appl. Mater. Interfaces* **2015**, *7* (13), 7206–7213. https://doi.org/10.1021/acsami.5b00292.

(35) Zuo, Z.; Liao, R.; Zhao, X.; Song, X.; Qiao, Z.; Guo, C.; Zhuang, A.; Yuan, Y. Anti-Frosting Performance of Superhydrophobic Surface with ZnO Nanorods. *Appl. Therm. Eng.* **2017**, *110*, 39–48. https://doi.org/10.1016/j.applthermaleng.2016.08.145.

(36) Zhao, G.; Zou, G.; Wang, W.; Geng, R.; Yan, X.; He, Z.; Liu, L.; Zhou, X.; Lv, J.; Wang, J. Competing Effects between Condensation and Self-Removal of Water Droplets Determine Antifrosting Performance of Superhydrophobic Surfaces. *ACS Appl. Mater. Interfaces* **2020**, *12* (6), 7805–7814. https://doi.org/10.1021/acsami.9b21704.

(37) Boreyko, J. B.; Hansen, R. R.; Murphy, K. R.; Nath, S.; Retterer, S. T.; Collier, C. P. Controlling Condensation and Frost Growth with Chemical Micropatterns. *Sci. Rep.* **2016**, *6*. https://doi.org/10.1038/srep19131.

(38) Ahmadi, S. F.; Nath, S.; Iliff, G. J.; Srijanto, B. R.; Collier, C. P.; Yue, P.; Boreyko, J. B. Passive Antifrosting Surfaces Using Microscopic Ice Patterns. *ACS Appl. Mater. Interfaces* **2018**, *10* (38), 32874–32884. https://doi.org/10.1021/acsami.8b11285.

(39) Novo, N. G. Di; Bagolini, A.; Pugno, N. M. Single Condensation Droplet Self-Ejection from Divergent Structures with Uniform Wettability. (Preprint) https://doi.org/10.48550/arXiv.2306.17635. submitted: July 2023.

(40) Saadi, N. S.; Hassan, L. B.; Karabacak, T. Metal Oxide Nanostructures by a Simple Hot Water Treatment. *Sci. Rep.* **2017**, No. July, 1–8. https://doi.org/10.1038/s41598-017-07783-8.

(41) Vermilyea, W. V. D. A. Aluminum + Water Reaction. *Trans. Faraday Soc* **1969**, *65*, 561–584.

(42) D. A. Vermilyea; W. Vedder. Inhibition of the Aluminum + Water Reaction. *Trans. Faraday Soc.* **1970**, *66*, 2644–2654.

(43) Crundwell, F. K. On the Mechanism of the Dissolution of Quartz and Silica in Aqueous Solutions. *ACS Omega* **2017**, *2* (3), 1116–1127. https://doi.org/10.1021/acsomega.7b00019.

(44) Arganda-Carreras, I.; Kaynig, V.; Rueden, C.; Eliceiri, K. W.; Schindelin, J.; Cardona, A.; Seung, H. S. Trainable Weka Segmentation: A Machine Learning Tool for Microscopy Pixel Classification. *Bioinformatics* **2017**, *33* (15), 2424–2426. https://doi.org/10.1093/bioinformatics/btx180.

(45) Haeri, M.; Haeri, M. ImageJ Plugin for Analysis of Porous Scaffolds Used in Tissue Engineering. *J. Open Res. Softw.* **2015**, *3*, 2–5. https://doi.org/10.5334/jors.bn.

(46) Seo, D.; Oh, S.; Moon, B.; Kim, H.; Kim, J.; Lee, C.; Nam, Y. Influence of Lubricant-Mediated Droplet Coalescence on Frosting Delay on Lubricant Impregnated Surfaces. *Int. J. Heat Mass Transf.* **2019**, *128*, 217–228. https://doi.org/10.1016/j.ijheatmasstransfer.2018.08.131.

(47) Shen, Y.; Zou, H.; Wang, S. Condensation Frosting on Micropillar Surfaces - Effect of Microscale Roughness on Ice Propagation. *Langmuir* **2020**, *36* (45), 13563–13574. https://doi.org/10.1021/acs.langmuir.0c02353.

(48) Hauer, L.; Wong, W. S. Y.; Sharifi-Aghili, A.; Kondic, L.; Vollmer, D. Frost Spreading and Pattern Formation on Microstructured Surfaces. *Phys. Rev. E* **2021**, *104* (4), 1–7. https://doi.org/10.1103/PhysRevE.104.044901.

(49) Jung, S.; Tiwari, M. K.; Doan, N. V.; Poulikakos, D. Mechanism of Supercooled Droplet Freezing on Surfaces. *Nat. Commun.* **2012**, *3*. https://doi.org/10.1038/ncomms1630.

(50) Kenneth G. Libbrecht. *Snow Crystals*; 2021.